\documentclass[12pt,letterpaper]{article}
\usepackage{authblk}
\usepackage{graphicx}
\usepackage{cite}

\let\oldabstract\abstract
\let\oldendabstract\endabstract
\makeatletter
\renewenvironment{abstract}
{%
               {\list{}{\addtolength{\leftmargin}{-2.5em} 
                        \listparindent 0em%
                        \itemindent    \listparindent%
                        \rightmargin   \leftmargin%
                        \parsep        \z@ \@plus\p@}%
                \item\relax}%
               {\endlist}%
\oldabstract}
{\oldendabstract}
\makeatother

\begin{document}

\title{Error detection using pneumatic logic}

\author[a]{Shane Hoang}
\author[a]{Mabel Shehada}
\author[a]{Zinal Patel}
\author[a]{Minh-Huy Tran}
\author[b]{Konstantinos Karydis}
\author[c]{Philip Brisk}
\author[a]{William H. Grover\footnote{Corresponding author; {\tt wgrover@engr.ucr.edu}
}}

\affil[a]{\small Department of Bioengineering, University of California, Riverside, 900 University Ave., Riverside, CA 92521}
\affil[b]{Department of Electrical and Computer Engineering, University of California, Riverside, 900 University Ave., Riverside, CA 92521}
\affil[c]{Department of Computer Science and Engineering, University of California, Riverside, 900 University Ave., Riverside, CA 92521}



\date{}
\maketitle

\begin{abstract}
Pneumatic systems are common in manufacturing, healthcare, transportation, robotics, and many other fields.  Failures in these systems can have very serious consequences, particularly if they go undetected.  In this work, we present an air-powered error detector device that can detect and respond to failures in pneumatically actuated systems.  The device contains 21 monolithic membrane valves that act like transistors in a pneumatic logic ``circuit'' that uses vacuum to represent TRUE and atmospheric pressure as FALSE.  Three pneumatic exclusive-OR (XOR) gates are used to calculate the parity bit corresponding to the values of several control bits.  If the calculated value of the parity bit differs from the expected value, then an error (like a leak or a blocked air line) has been detected and the device outputs a pneumatic error signal which can in turn be used to alert a user, shut down the system, or take some other action.  As a proof-of-concept, we used our pneumatic error detector to monitor the operation of a medical device, an intermittent pneumatic compression (IPC) device commonly used to prevent the formation of life-threatening blood clots in the wearer's legs.  Experiments confirm that when the IPC device was damaged, the pneumatic error detector immediately recognized the error (a leak) and alerted the wearer using sound.  By providing a simple and low-cost way to add fault detection to pneumatic actuation systems without using sensors, our pneumatic error detector can promote safety and reliability across the wide range of pneumatic systems.
\end{abstract}

\section*{Introduction}

Pneumatics are used in a wide variety of mechanical systems.  Many pneumatically actuated systems find applications in healthcare, manufacturing, transportation, robotics, and other areas where failures can have very serious consequences.  In these critical applications, it is desirable to endow these systems with error-detection strategies that can detect failures in the pneumatic actuation system (for example, a leak or a blockage) and take appropriate action (raise an alarm, shut down the system safely, and so on).  Current error-detection strategies employ electronic sensors that monitor air pressure or flow rate at various points in a system and relay this information to a separate control system (often a computer or microcontroller) for analysis and error mitigation.  This electronic monitoring hardware adds considerable complexity, size, and cost to the overall system.  This approach is also particularly problematic in smaller-scale systems such as soft robotics which use pneumatics to control air-filled actuators (e.g., \cite{zhang2018, janghorban2022, zhao2021, shao2022, kokkoni2020, shi2022, liu2020}). These robots are particularly sensitive to size, weight, and power (SWaP) considerations.  Considering that each independent actuator typically already has a separate pneumatic control line, adding yet another set of components for error detection further impacts SWaP efficiency and defeats many of the advantages of soft robotics (simplicity, autonomy, low cost, biomimetic design, and few or no electronic components).

In this work, we show that pneumatic logic can be used to detect errors in pneumatic systems without employing sensors. To achieve this, we use {\em monolithic membrane valves}, a microfluidic valving technology that was originally developed for controlling fluid flow in microfluidic ``lab on a chip'' devices \cite{grover2003} but was later adapted to create air-powered logic ``circuits'' for controlling microfluidic chips \cite{jensen2007, grover2006, duncan2013, jensen2010, jensen2010a, jensen2013, kim2013, linshiz2016, nguyen2012, duncan2015, ahrar2023} and more recently adapted for controlling soft robots \cite{hoang2021} and biomedical devices\cite{hoang2024}.  While a number of different approaches to pneumatic logic in soft robots have been demonstrated \cite{rothemund2018, drotman2021, lee2022, zhai2023}, monolithic membrane valves are particularly suitable for arranging in large numbers to form complex logical circuits, so we chose to use these valves to construct our pneumatic error detector.

There are many different methods for error detection in computing and communication systems.  In this work, we used {\em parity bits} for error detection; this fundamental yet effective error detection technique has been used in electronic computing since at least the early 1950s \cite{lukolf1952}.  In parity-bit-based error detection, the current values ({\bf 1} or {\bf 0}) of several binary bits are used to calculate the value of a parity bit.  For example, consider three binary bits with the values {\bf 1}, {\bf 1}, {\bf 0}.  To calculate the parity bit that corresponds to these three bits' values, we can use any one of these (mathematically equivalent) methods:
\begin{itemize}
\item Calculate the Boolean Exclusive OR (or XOR) of all the bits:  {\bf 1} XOR {\bf 1} XOR {\bf 0} = {\bf 0}.
\item Calculate the sum of the bits {\em modulo} 2:  {\bf 1} + {\bf 1} + {\bf 0} = 2 (mod 2) = {\bf 0}.
\item Count the number of 1's in the values of the bits; the parity bit is {\bf 1} if the count is an odd number and {\bf 0} if the count is an even number.  Since there are two 1's in {\bf 1}, {\bf 1}, {\bf 0} and two is even, the parity bit is {\bf 0}.
\end{itemize}
This ``expected'' parity bit value is then transmitted along with the original bits to some recipient, and the recipient repeats the parity bit calculation using the values of the bits they received.  If the bits' values were unchanged during transmission, then the parity bit's value will also be unchanged, and the recipient can be confident that no single-bit errors occurred during transmission.  However, if one bit changed state during transmission (for example, if {\bf 1}, {\bf 1}, {\bf 0} was received as {\bf 1}, {\bf 0}, {\bf 0}), then the parity bit calculated by the recipient would also change (in this case, from {\bf 0} to {\bf 1}) and would no longer match the expected parity bit.  The recipient would know that an error occurred and one of the received bit values is wrong.

Our pneumatic logic error detector uses air flowing through a network of 21 monolithic membrane valves to calculate a the value of a parity bit corresponding to the states of three pneumatic control signals.  If the calculated and expected parity bits differ at any point, then an error has been detected (one of the control signals is in the wrong state).  When this happens, the pneumatic error detector automatically outputs a pneumatic signal that can be used to alert a user, shut down the system, or take other action.  As a proof-of-concept, we used our pneumatic error detector to automatically detect different types of failures in the operation of an important medical device, an intermittent pneumatic compression (IPC) device that prevents the formation of life-threatening blood clots in the wearer's legs \cite{partsch2008, morris2008, chen2001}.  When a failure occurs (like an air leak or a blocked air line) that would compromise the efficacy of the IPC device and possibly endanger the wearer of the device, our pneumatic error detector senses this error and, in this demonstration, alerts the wearer or nearby healthcare professionals by blowing a whistle.  This pneumatic error detector is a direct and low-cost way to add error detection to a wide variety of pneumatic-controlled systems.

\section*{Results}

\begin{figure}[htbp]
\centering
\includegraphics[width=\linewidth]{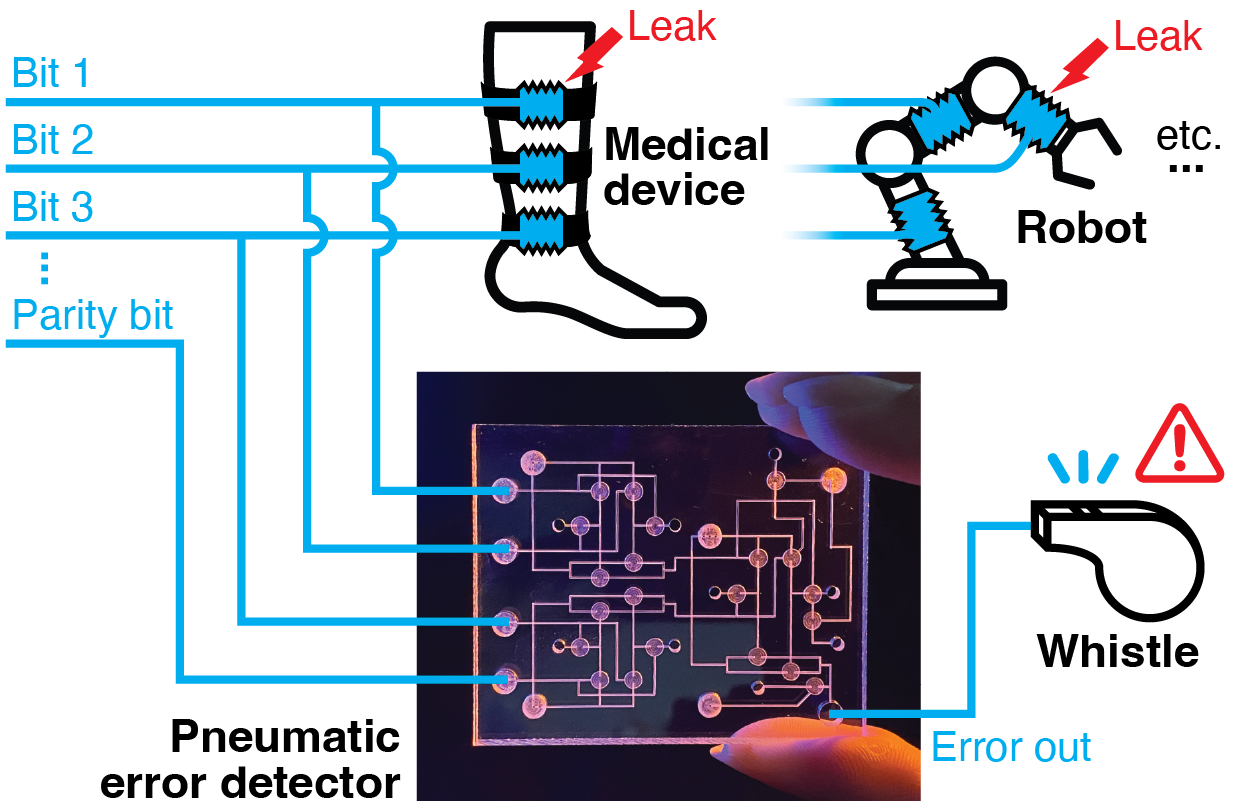}
\caption{Using our pneumatic error detector to detect problems during the operation of a typical pneumatically actuated system.  In this example, three pneumatic control lines (bits 1, 2, and 3) apply vacuum or atmospheric pressure to the system being controlled (a medical device, robot, etc.).  A fourth pneumatic control line contains a pneumatic signal that represents the parity bit corresponding to the values of the three control bits at each step during the operation sequence, with vacuum = {\bf 1} (True) and atmospheric pressure = {\bf 0} (False).  The pneumatic control and parity bit lines are connected to the pneumatic error detector, which uses an air-powered logic circuit consisting of 21 monolithic membrane valves to calculate the parity bit corresponding to the values of the control bits and compare it to the expected parity bit value.  If the two values for the parity bit are different, this indicates that one of the pneumatic signals is incorrect due to {\em e.g.}\ a leak occurring in the medical device or soft robot, and the error detector responds by automatically outputting {\bf 1} (vacuum) on an error line.  This pneumatic error signal can be used to alert the operator (using a whistle here), initiate a system shutdown, or take some other corrective action.}
\label{overview}
\end{figure}

Figure \ref{overview} provides an overview of using the pneumatic error detector to sense problems with the operation of a pneumatically actuated system.  The system is controlled by conventional electronic hardware:  a computer, microcontroller, programmable logic controller (PLC), field programmable gate array (FPGA), etc.  A program running on the electronic hardware controls the states---either {\bf 1} (True) or {\bf 0} (False)---of each of several control bits (in this example, control bits 1,  2, and  3).  These bits in turn control three solenoid valves which apply the requested pneumatic signals (vacuum for {\bf 1}, and atmospheric pressure for {\bf 0}) to the pneumatic system to be controlled.  This setup is representative of many pneumatic control systems that switch multiple independent pneumatic control lines between two different pressure states (like an IPC medical device on a patient's leg, an air-powered robot, and many other systems).

To add our pneumatic error detector to the control system shown at the top of Figure \ref{overview}, the program running on the electronic control hardware is modified to calculate the value of the parity bit that corresponds to the values of control bits 1, 2, and 3 at each step during the device operation sequence.  An additional solenoid valve is used to convert the parity bit calculated by the electronic hardware into its pneumatic representation (again using vacuum for {\bf 1} and atmospheric pressure for {\bf 0}). Next, near the system being controlled, the pneumatic control signals (bits 1, 2, and 3) and pneumatic parity bit are connected to the inputs of the pneumatic error detector.  The control signal connections are made using tee connectors so that the control signals still pass on to the system being controlled.  Finally, outputs from five additional solenoid valves are connected to the error detector using tubing; these outputs are used to power the error detector (using vacuum) and reset the error detector between operations (details in {\em Materials and Methods} below).

Using monolithic membrane valves and flowing air, the error detector repeats the parity bit calculation originally performed by the electronic control hardware and compares the resulting parity bit value to the one calculated by the electronic hardware. If the two parity bits agree (both are {\bf 0}, or both are {\bf 1}), then the values for control bits 1, 2, and 3 have passed successfully from the computer to the system being controlled---no error has occurred, and the error detector outputs {\bf 0} (atmospheric pressure).  However, if the two parity bits disagree (one is {\bf 0} and the other is {\bf 1}), then one of the pneumatic signals is different from what the computer intended---an error has occurred.  The error detector outputs {\bf 1} (vacuum) which, in this example, causes a whistle alarm to sound, alerting those nearby that an error has occurred.  In this manner, our pneumatic error detector can give pneumatically controlled systems the ability to detect and respond to errors in their own operation.

\subsection*{Pneumatic error detector design and operation}

\begin{figure}[htbp]
\centering
\includegraphics[width=0.9\linewidth]{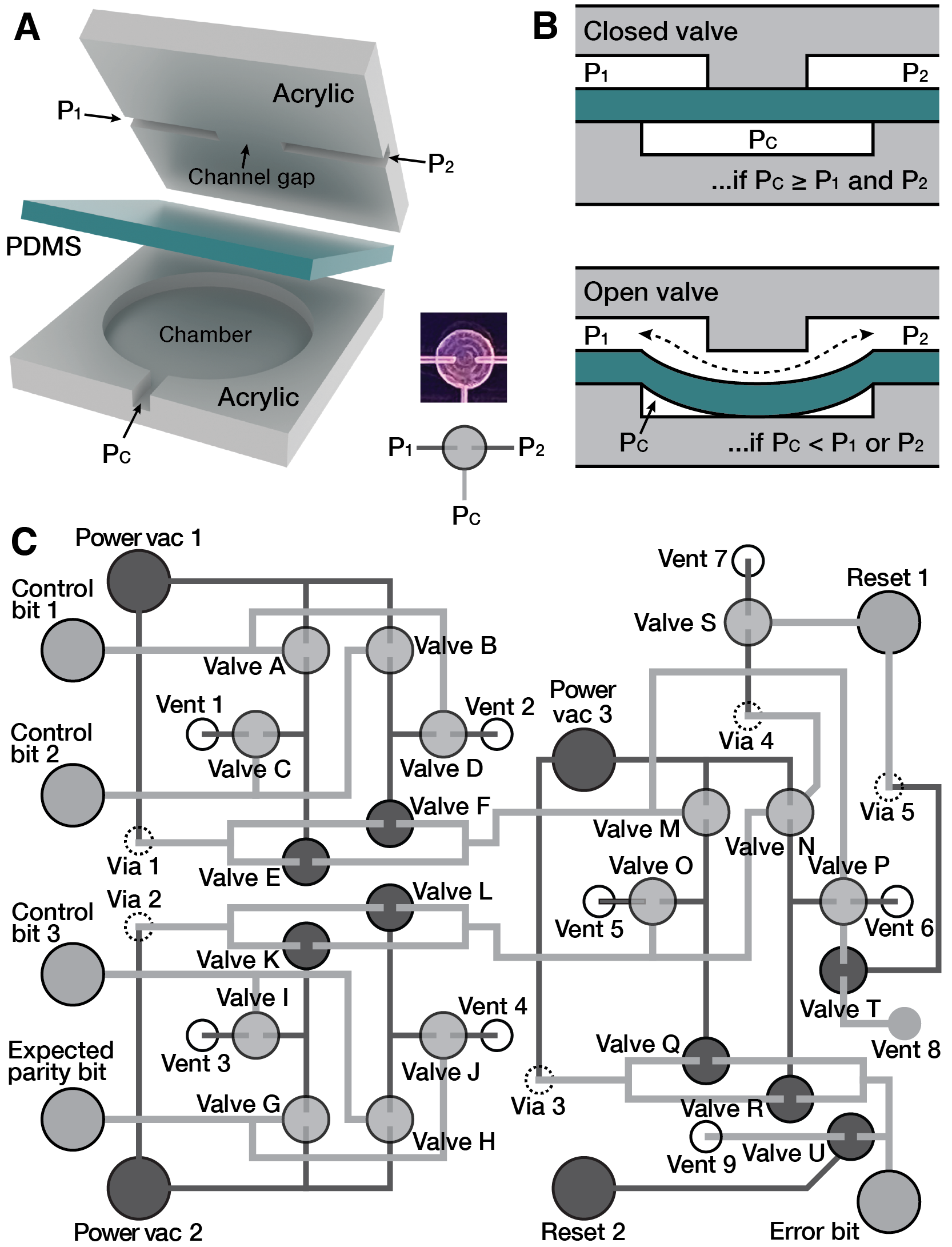}
\caption{Exploded (A) and cross-section (B) views of a single monolithic membrane valve.  The pneumatic error detector (C) contains 21 valves; it calculates the parity bit corresponding to three pneumatic control bit signals and compares the result to the input expected parity bit.  If the two values differ, then an error has been detected and the error bit output becomes {\bf 1} (vacuum); otherwise it outputs {\bf 0} (atmospheric pressure).}
\label{chip}
\end{figure}

Our pneumatic error detector consists of three layers:  a featureless polydimethylsiloxane (PDMS) silicone rubber membrane sandwiched between two engraved acrylic plastic sheets.  Monolithic membrane valves \cite{grover2003} are formed wherever a gap in an engraved channel in one acrylic layer is located directly across the PDMS membrane from an engraved chamber in the other acrylic layer, as shown in the exploded view in Figure \ref{chip}A.  A cross-section through a valve (Figure \ref{chip}B) shows that these valves are normally closed; the PDMS membrane normally rests against the channel gap and stops air from flowing across the gap.  When a vacuum is applied to the chamber, the PDMS membrane is pulled into the chamber and away from the channel gap; this creates a path for air to flow across the gap and the valve opens.  More generally, we can say that for a valve with pressures $P_1$ and $P_2$ at the two connections to the valved channel and pressure $P_C$ at the chamber:
\begin{itemize}
\item If $P_C \geq P_1$ and $P_C \geq P_2$, then the valve will be closed.
\item If $P_C < P_1$ or $P_C < P_2$, then the valve will be open; air will flow from channel 1 to channel 2 as long as $P_1 > P_2$, or from channel 2 to channel 1 as long as $P_2 > P_1$.
\end{itemize}

Multiple monolithic membrane valves can be connected together to form more complex pneumatic logic gates.  For example, two valves in series function as a Boolean AND gate:  air can flow through the valves only if valve A AND valve B both receive vacuum (that is, if A = {\bf 1} AND B = {\bf 1}).  Likewise, two valves in parallel function as a Boolean OR gate:  air can flow through the gate if either valve A OR valve B (or both) receive vacuum (in other words, if A = {\bf 1} OR B = {\bf 1}).  Finally, six valves and two vents (drilled holes to the atmosphere) can function as an exclusive-OR (XOR):  air flows through the valves if only Valve A receives vacuum, or if only Valve B receives vacuum, but not if both (or neither) receives vacuum.  Details on these and other valve-based pneumatic logic gates are available elsewhere \cite{jensen2007}.

\begin{figure}[htbp]
\centering
\includegraphics[width=0.95\linewidth]{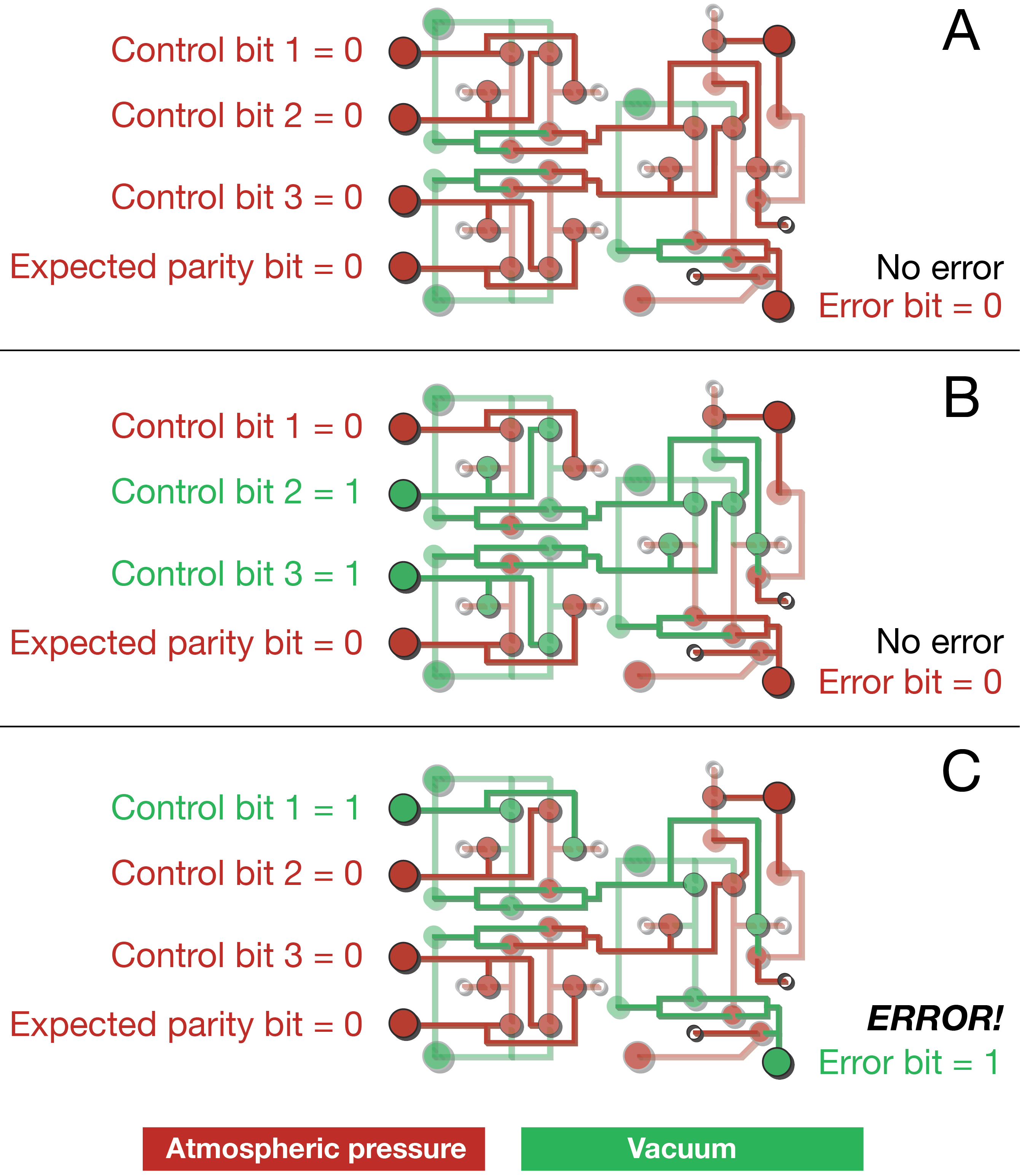}
\caption{Pressures inside the pneumatic error detector's channels (red for atmospheric pressure or {\bf 0}, and green for vacuum or {\bf 1}) during three example calculations.  In examples A and B, the error detector confirms that the expected and calculated parity bits match, so no error is detected and the error output remains at atmospheric pressure ({\bf 0}).  In the third example (C), the expected parity bit of {\bf 0} does not match the calculated parity bit of {\bf 1}, so the error detector outputs a vacuum ({\bf 1}) indicating a problem has been detected.}
\label{operation}
\end{figure}

The design of our pneumatic error detector is shown in Figures \ref{chip}D and E.  This pneumatic circuit comprises 21 valves: 18 of the valves are used in three XOR gates (valves A through F,  G through L, and M through R); and three additional valves (S through U) are used to ``unlatch'' or vent trapped vacuums to reset the device between operations.  The device has three control bit inputs (bits 1, 2, and 3), one expected parity bit input, one error bit output, three ``power vacuum'' inputs that receive vacuum to power the device, and two ``reset'' inputs that receive vacuum for resetting the device following operation.  The device also contains five ``vias'' (holes punched in the PDMS membrane before device assembly) to allow pneumatic signals to pass from one layer to another, and nine drilled vents to admit atmospheric-pressure air into the device.  Figure \ref{operation} depicts the contents (vacuum or atmospheric pressure) of every feature inside the device during three sample computations.

\subsection*{Testing the pneumatic error detector}

\begin{figure}[htbp]
\centering
\includegraphics[width=\linewidth]{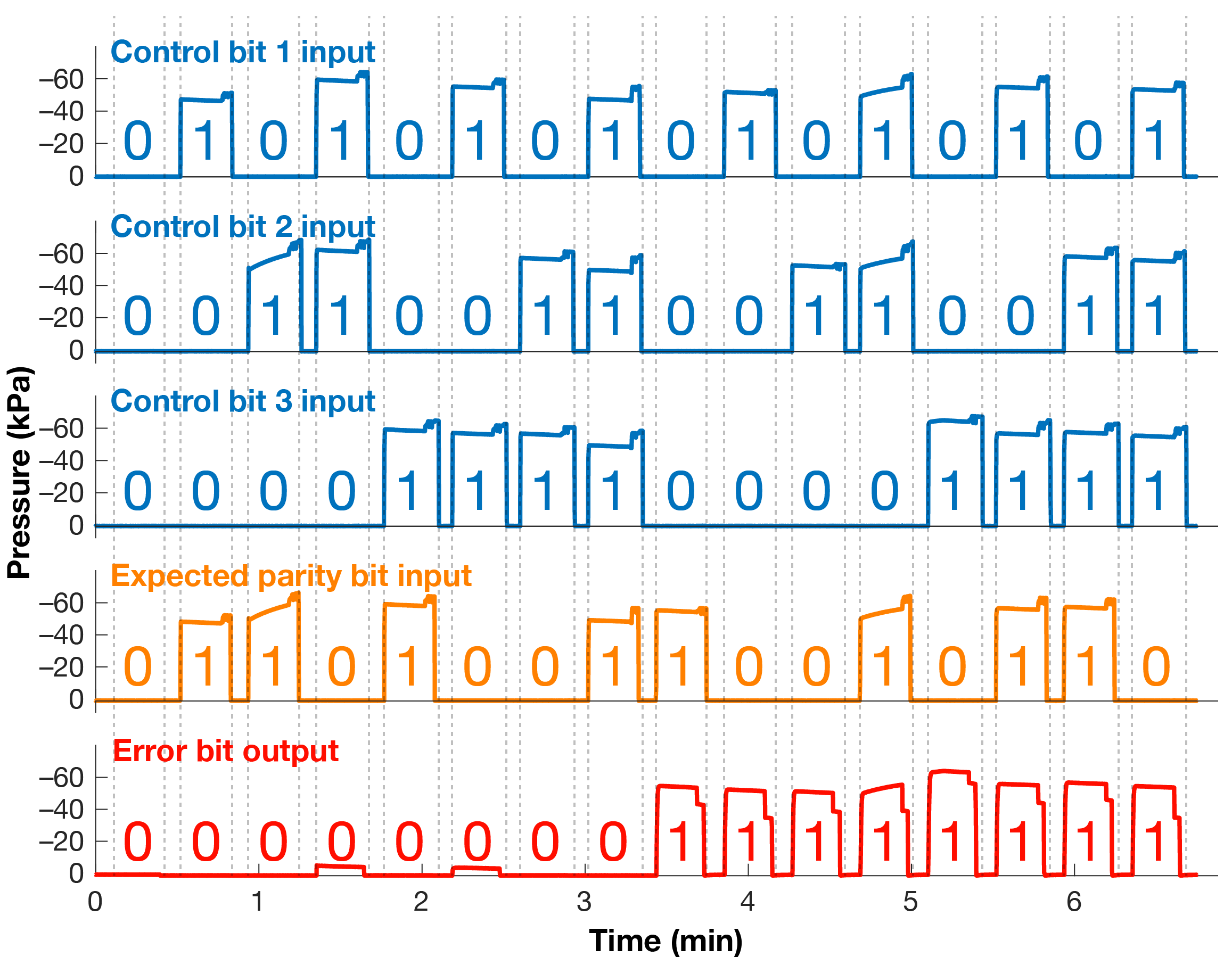}
\caption{Pressure measured at each of the three control bit inputs (blue), one expected parity bit input (orange), and one error output (red) while applying all 16 possible combinations of {\bf 1}'s (vacuums) and {\bf 0}'s (pressures) to the control bit inputs and parity bit input.  During the first eight combinations (times from 0 to 3.5 minutes), the expected parity bit is correct or consistent with the values of the three control bits, and the near-zero (atmospheric) pressures measured at the error output confirm that no error has occurred.  However, during the last eight combinations (times from 3.5 minutes to 7 minutes), the expected parity bit is intentionally incorrect (the opposite of what it should be), and the vacuum pressures measured at the error output confirm that the device has successfully detected these errors.  Results from additional experiments like this are provided in online {\em Supplementary Materials}.}
\label{states}
\end{figure}

To test the operation of our pneumatic error detector, we operated the device using all 16 possible combinations of {\bf 1}'s (vacuum) and {\bf 0}'s (atmospheric pressure) to the three control bit inputs and one expected parity bit input while measuring the pressure at the error output. Figure \ref{states} plots the pressure measured at each of the four inputs and one output during a typical experiment.  On the left half of Figure \ref{states}, the value of the expected parity bit is always correct (consistent with the provided values for the control bits), and the pressure measured at the error output remains at or close to atmospheric pressure ({\bf 0}), successfully indicating that no error has occurred.  However, in the right half of Figure \ref{states}, the value of the expected parity bit is intentionally incorrect, and the pressure measured at the error output always goes to vacuum, showing that each simulated error has been successfully detected.  In this particular experiment, we maintained each of the 16 possible assignments of {\bf 1}'s and {\bf 0}'s for 15 seconds; data from replicates of this experiment (as well as similar experiments that held each combination for 10 seconds and 5 seconds) are provided in online {\em Supplementary Materials}.

\subsection*{Sample case study:  Detecting errors in a model medical device}

\begin{figure*}[htbpp]
\centering
\includegraphics[width=\linewidth]{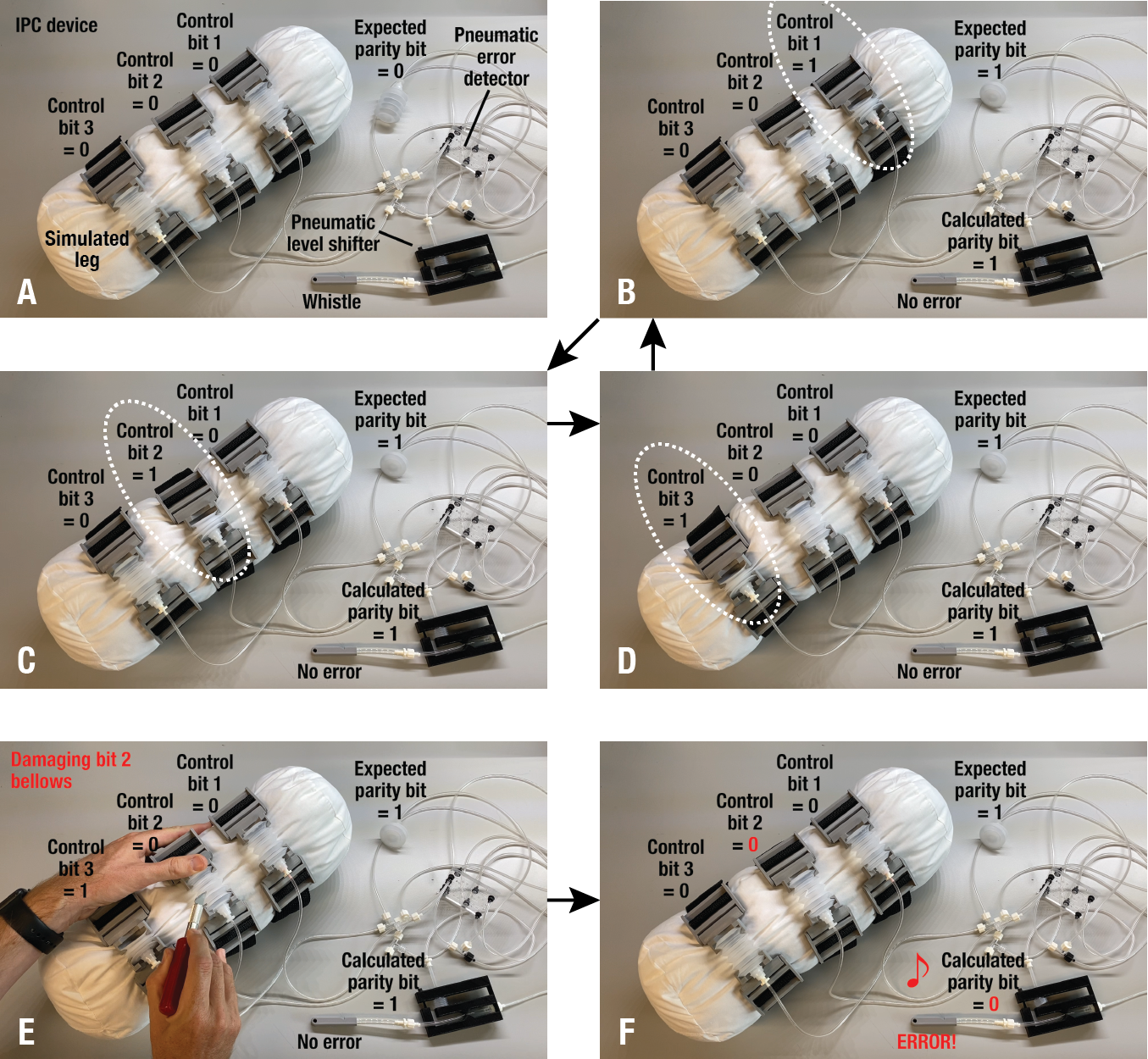}
\caption{Frames from a video recording (available as online {\em Supplementary Materials}) of the pneumatic error detector monitoring the operation of a model soft-robotic medical device, an intermittent pneumatic compression or IPC device used to prevent blood clots in the wearer's legs.  During normal operation (B $\rightarrow$ C $\rightarrow$ D $\rightarrow$ B $\rightarrow$ C $\rightarrow$ D...), the device control system contacts one bellows at a time and no errors are detected.  However, when a bellows is punctured to create a leak (E), the pneumatic error detector recognizes the mismatch between the expected ({\bf 1}) and calculated ({\bf 0}) parity bit values and automatically alerts the wearer by blowing a whistle (F). Detailed explanations of each frame are in the main text. }
\label{ipc}
\end{figure*}

To validate the pneumatic error detector in a real-world application, we used it to monitor a model soft-robotic medical device, an intermittent pneumatic compression (IPC) device commonly used to prevent the formation of blood clots in the wearer's legs.  Our model IPC device shown in Figure \ref{ipc}A consists of three flexible plastic bellows connected via 3D-printed buckles to nylon straps that wrap around a simulated leg.  When vacuum is applied to one of the IPC device's bellows, it contracts and squeezes the corresponding region on the simulated leg.  A computer program written in our valve control language {\em OCW} \cite{ocw} sets the values of the three control bits, which in turn control the three solenoid valves that apply vacuum (when the control bit is {\bf 1}) or atmospheric pressure (when the control bit is {\bf 0}) to the three IPC bellows.  The program contracts the bellows one-at-a-time in sequence:  first setting control bit 1 to {\bf 1} (shown in Figure \ref{ipc}B; the contracted bellows is indicated using a white dotted line), then setting control bit 2 to {\bf 1} (Figure  \ref{ipc}C), then setting control bit 3 to {\bf 1} (Figure \ref{ipc}D).  This pattern repeats over and over (Figure \ref{ipc}B $\rightarrow$ C $\rightarrow$ D $\rightarrow$ B $\rightarrow$ C $\rightarrow$ D...), creating a peristaltic squeezing motion that is meant to encourage blood flow in the leg.  The same computer program also calculates the value of the parity bit corresponding to the values of the three control bits at each step in the actuation pattern, and an additional solenoid valve outputs the pneumatic version of this expected parity bit ({\bf 1} = vacuum and {\bf 0} = atmospheric pressure) whenever error checking is desired.  The computer program also controls three solenoid valves that provide vacuum to power the error detector, and two solenoid valves that reset the error detector after operation.  An additional free bellows (labeled ``Expected parity bit'' in Figure \ref{ipc}) was added to the expected parity bit pneumatic line so that the state of this line can be visualized during operation (contracted bellows = {\bf 1} and extended bellows = {\bf 0}).  The three pneumatic control bit signals and one pneumatic expected parity bit signal are connected to the pneumatic error detector, which repeats the parity bit calculation on the three control signals and compares the result to the expected parity bit.  If the two values disagree, the error detector sets its error output to {\bf 1}  (vacuum).

In this demonstration, we wanted the pneumatic error detector to alert the wearer by blowing a whistle when an error is detected.  Since most whistles use positive pressure (not vacuum) to generate a sound, we needed a simple method for converting the vacuum at the error output to a positive pressure for powering the whistle.  We accomplished this by using a ``pneumatic level shifter'' developed as part of another project.  This level shifter (shown in Figure \ref{ipc}) consists of a small flexible plastic bellows mounted in a 3D-printed plastic frame. The bellows' motion is mechanically relayed to a pinch point through which runs tubing attached to a pressurized air supply.  When no error is detected by the attached error detector, the level shifter's bellows is at atmospheric pressure and is fully extended, holding the pinch point closed and blocking the flow of pressurized air in the tubing.  However, when an error is detected, the level shifter's bellows receives vacuum from the error detector and contracts; this opens the pinch point and allows pressurized air to flow through the tubing and into the attached whistle, which makes a sound and alerts the wearer of a problem.

We successfully demonstrated error detection under two different modes of operation for the IPC device.  In the first mode, the pneumatic error detector was operated after every change in the values of the control bits 1, 2, and 3.  This mode offers continuous error checking (detecting an error as early as possible), but this comes at the expense of overall speed (the pneumatic error detector takes about one second to operate and about five seconds to reset after operation, so in this mode the control bits can only be updated every few seconds).  A video recording of the IPC system in this mode of operation is available as online {\em Supplementary Materials}; key snapshots are shown in Figure \ref{ipc}.  Over several minutes of normal operation (repeated cycling through the states shown in Figure \ref{ipc}B, C, and D, operating the pneumatic error detector after each step), the error output remained at atmospheric pressure and the whistle remained silent; this is expected during normal error-free operation.  Then, in Figure \ref{ipc}E, we used a knife to puncture the IPC bellows connected to control bit 2.  The next time that the system attempted to contract the damaged bellows by setting control bit 2 to {\bf 1} (vacuum), the vacuum was exhausted through the puncture in the bellows, so the error detector sensed that control bit 2 was {\bf 0} (atmospheric pressure).  The error detector then used this value along with control bit 1 ({\bf 0}) and control bit 3 ({\bf 0}) to calculate a parity bit of {\bf 0} XOR {\bf 0} XOR {\bf 0} = {\bf 0}.  This differed from the value of the expected parity bit input ({\bf 1}), which caused the error detector to output {\bf 1} (vacuum) to indicate the error.  Finally, the pneumatic level shifter converted this signal to a positive pressure which caused the whistle to blow (Figure \ref{ipc}F; see also the video in online {\em Supplementary Materials}).  The whistle continued to sound every time that the error was detected again, until the leak was repaired.  In this manner, the error detector successfully detected damage to the IPC device mere seconds after the damage occurred and notified the wearer about the problem.

In the second mode of operation we demonstrated, the IPC device was alternated between two phases:  a ``run'' phase, during which the control bits can be changed at high speeds without activating the pneumatic error detector; and a ``check'' phase, during which the pneumatic error detector checks each control bit in turn. This mode of operation offers periods of much faster operation (the control bits can be updated several times per second during the ``run'' phase) at the expense of error checking frequency (errors are only detected during the ``check'' phase).  In the video recording in {\em Supplementary Materials}, the IPC device alternated between spending 22.5 seconds in the ``run'' phase (during which a bellows was actuated every 750 milliseconds) and 39 seconds in the ``check'' phase (during which the system applied vacuum to the control bits one-at-a-time while the pneumatic error detector checked for errors).  When the IPC device was damaged during the ``run'' phase by using scissors to cut the tubing leading to control bit 3's bellows, the pneumatic error detector successfully sensed this damage and blew the whistle 45 seconds later during the system's next ``check'' phase.  Finally, this video also demonstrates that the pneumatic error detector's error signal is automatically reset after fixing the error:  when we repaired the cut tubing, the whistle was again silent in subsequent ``check'' phases.

\section*{Discussion}

In this work we demonstrated that pneumatic logic can be used to detect failures in pneumatic systems. We conclude by discussing the implications, practical limitations, and future directions for this technology.

One noteworthy aspect of air-powered error detection is its low cost and simplicity.  The material cost of the pneumatic error detector shown in Figure \ref{chip}E is only \$0.93 USD, and the design is amenable to mass production.  While our current device does require some additional electromechanical control hardware (specifically, one additional solenoid valve for providing the pneumatic error detector with the expected value of the parity bit, and five additional solenoid valves for powering and resetting the pneumatic error detector), the device provides a sophisticated level of error detection capabilities without the need for any sensors or sensor-reading electronics (or any electronics whatsoever on the system being controlled).  This is particularly attractive for robotic applications in environments not suitable for electronics (for example, under water, around explosive or flammable materials, in high-radiation areas, in contact with humans \cite{kokkoni2020, sahin2022, mucchiani2022}, and so on)---in these applications the pneumatic error detector can detect failures in any of the robot's control lines with {\em no electronic hardware on the robot itself}.

Additionally, pneumatic logic circuits for error detection are not limited to the parity-bit-based error detection algorithm demonstrated here.  Parity bits were used to successfully detect all the errors demonstrated in this work, but they cannot detect all possible errors.  For example, if {\em two} control bits (or any other even number of control bits) had the wrong values {\em at the same time}, then the associated value of the parity bit would not change, and a parity-bit-based error detection like the one shown here would not be able to detect those simultaneous errors.  If multiple simultaneous errors are a realistic concern in an application, then pneumatic error detectors could be designed that use other error detection schemes.  For example, algorithms like cyclic redundancy checks \cite{peterson1961} and Fletcher's checksum \cite{fletcher1982} can detect multiple simultaneous errors.  And while these algorithms are more complex than the parity bit approach shown here, the recent demonstration of a {\em complete programmable computer} using monolithic membrane valve-based pneumatic logic \cite{ahrar2023} shows that even complex computations can be performed in pneumatic logic circuits.

Our pneumatic error detector runs on vacuum (pressures lower than atmospheric pressure).  Using vacuum allows us to use monolithic membrane valves \cite{grover2003} as the ``transistors'' in our circuits; these vacuum-operated normally-closed valves are generally far more amenable to use in complex pneumatic logic circuits \cite{jensen2007, grover2006, duncan2013, jensen2010, jensen2010a, jensen2013, kim2013, linshiz2016, nguyen2012, duncan2015, hoang2021, ahrar2023} than pressure-operated normally-open valves are.  If the pneumatic system to be monitored also runs on vacuum, then our pneumatic error detector can be connected directly to the system being controlled.  For systems that run on positive pressures (greater than atmospheric pressure), the simple 3D-printed ``pneumatic level shifter'' shown in Figure \ref{ipc} could be used to convert signals between pressure and vacuum as necessary (like a voltage level shifter in electronics); research on this topic is ongoing.

Finally, while this work focused primarily on a biomedical application for the pneumatic error detector, this represents only the ``tip of the iceberg'' of applications for this technology.  In principle, any pneumatically controlled system could gain sophisticated fault detection capabilities without sensors by adding a pneumatic logic circuit like our error detector.  This simple and low-cost approach to error detection can promote safety and reliability across a wide range of important application areas.

\section*{Supplementary Materials}

\begin{itemize}
\item Results from five replicates of the experiment shown in Figure \ref{states} for three different durations (5, 10, and 15 s) for each of the 16 possible states of the error detector.
\item Video recording of the pneumatic error detector identifying faults during the operation of our model intermittent pneumatic compression (IPC) device
\item CAD files containing the design of the pneumatic error detector
\item CAD files containing the design of the 3D-printed IPC device
\item Arduino and Python code and Eagle PCB design files for the multichannel pressure logger used to acquire data in Figure \ref{states}
\end{itemize}


\section*{Materials and methods}

\subsection*{Pneumatic error detector design and fabrication}

The design of the pneumatic error detector was created in Adobe Illustrator (file available as online {\em Supplementary Materials}) and exported as SVG files for milling into two acrylic substrates (each 6.35 cm wide by 5.08 cm high by 3 mm thick) using a desktop CNC mill (Bantam Tools, Peekskill, New York).  Channels were engraved at a width of 284 $\mu$m, and valve displacement chambers were engraved with a diameter of 3 mm; both features had a depth of 254 $\mu$m.  Vents (locations where atmospheric-pressure air can enter the device) were milled as through holes with diameters of 2 mm, and the error output port was milled as a through hole with 4 mm diameter.  The other input ports (the control bits and expected parity bit inputs, the vacuum inputs, and the ``unlatch'' reset inputs) were milled with two depths: 4 mm diameter for the first 2.75 mm, then 284 $\mu$m for the remaining 0.75 mm; this narrowing allows for air flow while preventing the polydimethylsiloxane (PDMS) membrane from being pulled into the input port by vacuum after the device is bonded.  The ports were then tapped with 10-32 threads.

To prepare them for bonding, the two pieces of acrylic were cleaned using 99.5\% isopropanol and soaked in a 5\% (volume/volume) solution of 3-aminopropyltriethoxysilane (Sigma-Aldrich, St. Louis, MO), diluted in purified water, for 20 minutes.  The pieces were then rinsed in purified water and blown dry using nitrogen.  Next, a 254 $\mu$m thick sheet of PDMS (HT-6240; Bisco Silicones/Rogers Corporation, Carol Stream, IL) was cut out to the size of the acrylic pieces and punched with via holes (locations where the pneumatic signal needs to cross from one side of the PDMS membrane to the other) using a 3 mm diameter biopsy punch (Electron Microscopy Sciences, Hatfield, PA).  The bonding surfaces of both the acrylic and PDMS layers were then treated with a handheld corona treater (BD-20AC; Electro-Technic Products, Chicago, IL) for 1 minute, then the acrylic and PDMS layers were assembled together to form the completed stack shown in Figure \ref{chip}.  The bonded device was clamped overnight to allow the bond to strengthen, then barbed tubing connectors were screwed into the ports.

\subsection*{Intermittent pneumatic compression system design and fabrication}

To test the pneumatic error detector with a model system representative of many soft robotic and medical applications, we designed and fabricated the model intermittent pneumatic compression (IPC) device shown in Figure \ref{ipc}.  Plastic bellows intended for dispensing applications (``Kitchen Witch'' cake decorating set; amazon.com) were connected via custom 3D-printed adapters (CAD files available as online {\em Supplementary Materials}) to nylon webbing straps that encircle a simulated leg made from a fabric-covered cylinder of polyester batting.  When vacuum is applied to one of the bellows, it contracts and squeezes the simulated leg.

\subsection*{Pneumatic control and measurement system}

A computer running LabVIEW and our {\em OCW} valve control software \cite{ocw} was used to control a bank of nine ``2 way, 3 ported'' solenoid valves (S070B-6BC, SMC Corporation of America; Noblesville, Indiana).  Three solenoid valves provided vacuum ($-68$ kPa) or atmospheric pressure (0 kPa) to the three bellows on the model IPC device as well as the three control bits on the pneumatic error detector (see Figure \ref{overview}).  One solenoid valve provided vacuum or atmospheric pressure to the expected parity bit input on the pneumatic error detector.  Three solenoid valves provided vacuum to the ``power'' vacuum inputs on the pneumatic error detector (see Figure \ref{chip}D).  Finally, two solenoid valves provided vacuum to the ``reset''  inputs on the pneumatic error detector.

To characterize the performance of the pneumatic error detector, a custom Arduino-based multichannel pressure sensor circuit utilizing differential pressure gauges (MPX4250DP; NXP Semiconductors, Austin, TX) and a data-logging Python computer program were used to monitor the pressures at the three control bit inputs, one expected parity bit input, and one error bit output (data shown in Figure \ref{states}).  The printed circuit board design and Arduino and Python code for the multichannel pressure monitor are available as {\em Supplementary Materials}.

\subsection*{Operating the pneumatic error detector}

The pneumatic error detector is connected in parallel with the pneumatic system being controlled using tee junctions, as illustrated in Figure \ref{overview} and photographed in Figure \ref{ipc}.  This ensures that the pneumatic signals in the system being controlled are also available to the error detector's control bit inputs.  Additionally, the pneumatic signal representing the current expected parity bit value is also connected to the pneumatic error detector.  

To operate the pneumatic error detector, the ``power'' vacuum solenoid valves apply vacuum to the detector's ``power'' inputs.  The pneumatic error detector then automatically calculates the value of the parity bit corresponding to the current values of the control bits, compares the calculated parity bit to the expected parity bit, and applies a vacuum to the error bit output if the two parity bit values disagree.  This error-indicating vacuum will continue for as long as the pneumatic error detector remains in this state. 

To reset the pneumatic error detector, first the ``power'' vacuum inputs are turned off in sequence (first power 3, then power 2, and finally power 1).  Next, vacuum is momentarily applied to the detector's ``reset'' input; this opens valves S, T, and U and vents regions in the device that may contain trapped or ``latched'' vacuum that could otherwise cause the device to malfunction during the next error detection cycle if not vented.  Finally, the control bits and expected parity bit are reset to {\bf 0} (atmospheric pressure).  This reset cycle takes about five seconds, after which the pneumatic error detector is again ready to detect errors.

For applications that are incompatible with the five-second reset cycle ({\em e.g.,} applications that require rapid actuation of the control bits), the system can alternate between ``run'' and ``check'' phases as described above.  During the ``run'' phase, the ``power'' vacuum inputs to the pneumatic error checker are kept off (at atmospheric pressure), so the control bits can be set to any desired pattern (and changed rapidly as desired) without activating the error detector.  During the ``check'' phase, the vacuum supplies to the pneumatic error detector's ``power'' inputs are turned on and the control system sets the control and expected parity bits to whatever pattern is needed to check for errors (for example, setting all control bits to {\bf 1} or vacuum, or setting each control bit to {\bf 1} in sequence).  If an error is detected, then the pneumatic error detector will output a vacuum.  Otherwise, once the ``check'' phase is completed and the error detector is reset, the system can reenter the ``run'' phase.  In this manner, the system can alternate between ``run'' and ``check'' phases with whatever frequency is suitable for a given application.

\subsection*{Acknowledgments}

This work was supported by the National Science Foundation's Division of Civil, Mechanical and Manufacturing Innovation (CMMI) under award numbers 1740052 (PB), 2046270 (KK), and 2133084 (KK); the National Science Foundation's Division of Computing and Communication Frontiers (CCF) under award number 2019362 (WHG); and the National Science Foundation's Division of Biological Infrastructure (DBI) under award number 2131428 (WHG); {\tt https://nsf.gov}.  The funders played no role in the study design, data collection and analysis, decision to publish, or preparation of the manuscript.

\bibliographystyle{wgrover}
\bibliography{parity}

\end{document}